# ESO White Paper: 'The nature of galactic spiral arms"

Robert J J Grand[1], Martin Roth[2]

[1]Astrophysics Research Institute, Liverpool John Moores University, UK
[2]Leibniz-Institut für Astrophysik Potsdam (AIP), Germany

# Rationale:

Spiral arms are the key features of spiral galaxies, but their nature and influence on galaxy evolution is not known; current simulations and observations present an unclear/mixed picture. There are two main groups of theories for spiral arms: the long-lived, rigidly rotating spiral density waves; and dynamic, transient and recurring spiral arms. A key test of these scenarios is the peculiar velocity patterns around the spiral arms. However, observations needed measure the kinematics of stars on the necessary sub-kpc scales are limited with current IFU facilities. Given that cosmological simulations are becoming detailed enough to provide meaningful predictions for the nature of spiral arms based on kinematics maps of galaxies, there is currently an observational hole that prohibits comparisons that may end a century-long debate on the theory of spiral arms and give unprecedented insight into their impact on long-term galaxy evolution.

# Context:

Spiral galaxies are responsible for the main mode of main sequence star formation throughout the Universe, and host some of the most beautiful galactic structures ever seen - spiral arms. Spiral arms are believed to play a significant role in the secular evolution of galaxies, for example, the radial mixing and migration of stars from their birth radii as a result of dynamical interaction with a bar and spiral arms (Freeman & Bland-Hawthorn 2002; Sellwood & Binney 2002). Not only may radial migration shape several galactic chemo-dynamic trends, such as the flat age-metallicity relation observed in the Milky Way, e.g., Casagrande et al. (2011); the chemically distinct Galactic discs, e.g., Schoenrich & Binney (2009); and U-shaped radial age/metallicity/colour profiles of galaxies, e.g., Roskar et al. (2008).

The prevalence and impact of spiral-induced migration is tied to the physical nature of spiral arms – a still debated and unsolved problem (e.g. Sellwood & Masters 2022). A key property of spiral arms is the pattern speed – the velocity at which they rotate around the galaxy as a function of radius – whose precise form dictates the longevity of spiral arms and the location of important dynamical resonances at which radial migration is very strong. In the well-known spiral density wave theory (e.g. Lin & Shu 1964), the pattern speed is independent of radius and therefore there is a single radius at which the spiral arms and stars rotate at the same speed – the co-rotation radius. At this radius, stars feel smooth and prolonged torques from the spirals and exchange angular momentum with them, which leads to stars moving inwards and outwards without gaining random motion. Inside (outside) of the co-rotation radius, stars and gas move faster (slower) than the spiral arms, which can lead to shocks in the gas as it enters the spirals (Roberts 1970). This leads to observational predictions of shock-induced star formation leaving a temporal-spatial gradient of young star-formation tracers in the tangential direction across spiral arms (e.g. Ferreras 2011).

If these density wave "modes" are long-lived, such the induced radial migration would put stars on trapped horseshoe orbits around the singular resonance; stars would not move away from resonances in the long-term. Sellwood & Binney (2002) showed that transient spiral modes are needed to make sure migration accumulates over time. Numerical simulations have often shown transient, recurrent types of spiral arms that disappear and reappear on dynamical timescales, with some simulations arguing for "dynamic" or "co-rotating" spiral arms (e.g. Grand et al. 2012ab; Baba et al. 2013), whereby the pattern speed closely matches the galactic rotation curve at all radii. Some of the main signatures of this scenario are peculiar streaming motions that can range from a few up to ~20 km/s along spiral arms: the velocity field is locally tangentially slow and outward moving on the trailing edges of spiral arms, and vice versa for the leading edges – this contrasts with density wave theory which predicts inward radial velocity fields on the locus of the spiral arm and outward motions between the arms. In addition, stars and gas on the trailing edges of spiral arms are predicted to be more metal-rich compared to stars on the leading edges at the same radius for galaxies that possess a negative radial metallicity gradient:

thus, the amplitude of this variation informs the radial metallicity profile as well as the strength of migration.

The vast majority of numerical studies of the nature of spiral structure are idealised N-body simulations of initially smooth exponential discs embedded in a spherical dark matter halo. These have predicted a range of spiral arm theories, from co-rotating to multiple modes to bar-induced manifolds. Numerical resolution has also been raised as an important factor; D'Onghia et al. 2013 highlighted that simulations with <1e8 disc particles (the majority of simulations) can spontaneously form spiral arms through numerical shot noise, which is hardly ideal. However, while these high-resolution simulations demonstrate a fascinating degree of non-linearity of spiral arm formation, such simulations are overly simplified in their physics and lack the realistic cosmological environment and associated spectrum of perturbations.

Over the past decade, cosmological simulation within the $\Lambda$CDM paradigm have advanced from a description of the large-scale structure of the Universe and its evolution merely based on Dark Matter to now include baryons, the cycle of matter, and chemical evolution within individual galaxies (e.g. Auriga; Grand et al. 2017; 2024, see Crain & van de Voort 2023 for a review). These simulations have provided testable observable predictions for spiral arms: for example, the study of Sanchez-Menguiano et al. (2016) found evidence for peculiar gas motions in the spiral galaxy NGC 6754 consistent with dynamic spiral arms. However, the limited resolution makes detailed predictions difficult to achieve. The recent Auriga Superstars (Grand et al. 2023; Pakmor et al. 2025) employs a novel approach to boost the resolution of star particles in massive spiral discs to ~ 800 Solar Masses - 1e8 star particles identified by D'Onghia et al. 2013 in the full cosmological environment for the first time. This will enable predictions for stellar chemo-kinematics at an unprecedented level of detail and accuracy, enabling comparison of observable predictions with star-by-star observations in external spiral galaxies, and as such a deeper understanding of how stellar orbits are affected by spiral arms to produce any peculiar/streaming motions seen in integrated velocity maps.

## **What is limiting progress?**

Observational tests for models describing the chemical and kinematic evolution of galaxies have initially relied on global spectroscopy e.g., from SDSS which, however, is limited by the selective spatial coverage of the aperture and the associated bias. Significant progress over this limitation has been achieved with integral field spectroscopy (IFS), and surveys like CALIFA (Sanchez et al. 2012), SAMI (Bryant et al. 2015), or MaNGA (Bundy et al. 2015), but still limited to physical areas on kpc scales, owing to the angular resolution of the IFS spaxels, and the distance of the galaxies.

The development of crowded field integral field spectroscopy (Kamann et al. 2013) has provided the tools to resolve galaxies out to distances of 2 Mpc and beyond at pc scales into
individual stars, HII regions, diffuse ionized gas (DIG), supernova remnants, planetary nebulae, etc., at unprecedented sensitivity. A first attempt with MUSE observations of the nearby galaxy NGC300 (d=1.88 Mpc) has demonstrated this capability, e.g., pilot study by Roth et al. (2018), study on BA-type supergiant stars by Gonzalez-Tora et al. (2022), faint HII regions by Micheva et al. (2022), and PNLF by Soemitro et al. (2023).

A review on crowded field IFS by Roth et al. (2019) provides an overview of the technique and further studies, e.g., R136 in 30Dor (Castro et al. 2018), Leo P (Evans et al. 2019), NGC300 (McLeod et al. 2020), and others. The study of massive stars in nearby galaxies in particular was identified as a highlight science case for BlueMUSE (Richard et al. 2019). However, a major limitation of MUSE/BlueMUSE is the relatively small field of view in comparison to the angular size of nearby galaxies. Unfortunately, this prohibits us from obtaining a holistic view of the full spiral arms and associated kinematic patterns needed to compare with predictions from simulations required to place constraints on the nature of spiral arms.

## What is needed?

IFS coverage of the full extent of face-on nearby galaxies, including the faint outskirts, at moderate spectral resolution with the additional capability to obtain medium or high-resolution spectra for pre-selected objects with a Multi Object Spectrograph. The simultaneous use of IFS and MOS will provide detailed chemo-kinematic maps on the basis of individual stars with a large field of view covering entire discs of galaxies up to distances of ∼ 8 Mpc. To measure the peculiar velocity kinematics required to constrain spiral arm dynamical influences, a precision of order 1 km/s would be highly desirable.

Galaxy targets could be selected to include those observed by other surveys such as the PHANGS survey, which has conducted an ALMA large programme for nearby galaxies with CO maps and HST & JWST imaging (e.g. Querejeta et al. 2025). This would provide a synergy with PHANGS that connects the spiral arm dynamics to the physics of star formation as gas falls into spiral arms, and the associated presence/absence of shocks and shearing motions.